\newcommand{\be}{\begin{equation}}\newcommand{\ee}{\end{equation}}
\newcommand{\bea}{\begin{eqnarray}}\newcommand{\eea}{\end{eqnarray}}
\newcommand{\nn}{\nonumber \\}\newcommand{\p}[1]{(\ref{#1})}
\documentstyle[12pt]{article}

\begin{document}
\renewcommand{\thefootnote}{\fnsymbol{footnote}}
\thispagestyle{empty}
{\hfill LNF-94/013 (P) }\vspace{0.5cm} \\
\begin{center}
{\large\bf NONLINEAR REALIZATIONS OF THE $W_3^{(2)}$ ALGEBRA}
\vspace{0.5cm} \\
S. Bellucci${}^a$\footnote{E-mail: bellucci@lnf.infn.it},
V. Gribanov${}^b$,
S. Krivonos${}^{a,c}$\footnote{E-mail: krivonos@thsun1.jinr.dubna.su}
and A. Pashnev${}^c$\footnote{E-mail: pashnev@thsun1.jinr.dubna.su}
\vspace{0.5cm} \\
${}^a${\it INFN - Laboratori Nazionali di Frascati,
P.O.Box 13, I-00044 Frascati, Italy}  \\
${}^b${\it Voronezh State University, Voronezh, Russia} \\
${}^c${\it JINR--Bogoliubov Theoretical Laboratory,
P.O.Box 79, 101 000 Moscow, Russia} \vspace{0.5cm} \\
{\bf Abstract}
\end{center}
In this letter
we consider the nonlinear realizations of the classical Polyakov's
algebra $W_3^{(2)}$. The coset space method and the covariant reduction
procedure allow us to deduce the Boussinesq equation with interchanged
space and evolution coordinates.
By adding one more space coordinate and
introducing two copies of the $W_3^{(2)}$ algebra, the same method yields the
$sl(3,R)$ Toda lattice equations.\vspace{1cm} \\

\renewcommand{\thefootnote}{\arabic{footnote}}
\setcounter{footnote}0
\setcounter{equation}0
\section{Introduction}
Recently it was realized [1-3] that the classical $W_N^{(l)}$
algebras,
considered as second Hamiltonian structures, give rise to the integrable
hierarchy resulting from the interchange of the $x$ and $t$ variables in the
$sl(N)$ KdV one.
In particular, the Boussinesq equation in $x$-evolution has been
constructed along this line \cite{r1} starting from the classical $W_3^{(2)}$
algebra \cite{Polyakov}.
In view of the growing interest in theories with fractional spins, it is
important to understand, from various points of view, the geometric origin
of these $W_N^{(l)}$ symmetries.

The idea of a geometric set-up for classical infinite-dimensional algebras
(covariant reduction method) goes
back to Ref.\cite{IK1} . There, it was shown  that the Liouville
equation is intimately related to the intrinsic geometry of the
Virasoro algebra:
it singles out a two-dimensional fully geodesic surface in the coset space
of the Virasoro group. The extension of this method to the case of nonlinear
$W_N$-type algebras has been proposed in \cite{IKP} ,
and $sl_3$-Toda \cite{IKP}
and Boussinesq equation \cite{IKM}
have been constructed starting from the $W_3$ algebra.
In this letter we apply
the covariant reduction method to the simplest case
of $W_3^{(2)}$ algebra \cite{Polyakov} , i.e. the
first representative of the nonlinear algebras with
generators having non-canonical spins.

This letter is organized as follows.
In Sec. 2, the coset space method and the covariant reduction
procedure, applied to the  $W_3^{(2)}$ algebra,
allow us to deduce the Boussinesq equation in $x$-evolution together with
the Miura maps.
In Sec. 3, by adding one more space coordinate and
introducing two copies of the $W_3^{(2)}$ algebra,
the same method yields the $sl_3$ Toda lattice equations.
Finally, the conclusions of this work are drawn in Sec. 4.

\setcounter{equation}0
\section{Nonlinear Realizations of $W_3^{(2)}$ and
 $x$-Boussinesq equation}
The main idea of extending the nonlinear realization method to the
nonlinear
$W$-type algebras\cite{IKP} implies replacing the latter  by the linear
algebras
$W^{\infty}$ and then constructing the coset realizations of these
symmetries. The original $W$ symmetry appears as a particular field
realization of $W^{\infty}$. In this Section we construct  a set of
nonlinear realizations of $W_3^{(2)\infty}$ and show that these
realizations have a deep relation to the $x$-Boussinesq equation and
Miura maps for it.

We start from the commutation relations of the
$W_3^{(2)}$ algebra \cite{Polyakov}
\begin{eqnarray} \label{1}
\left[ L_n , L_m \right] & = & (n - m) L_{n+m}+ \frac{c}{12}
(n^3 -n) \delta_{n+m} \; ,\nn
\left[ L_n , G_r^{\pm}\right] & = &
   \left( \frac{n}{2}-r \right) \; G_{n+r}^{\pm} \; , \;
\left[ L_n , U_m \right]  = - m \; U_{n+m} \; , \nn
\left[ U_n , G_r^{\pm}\right]  & = & \pm G_{n+r}^{\pm} \; , \;
\left[ U_n , U_m \right]  =  -\frac{c}{9} n \delta_{n+m} \; , \nn
\left[ G_r^{+} , G_s^{-} \right]
& = & -\frac{c}{6}(r^2-\frac{1}{4})\delta_{r+s}
+\frac{3}{2}(r-s)U_{r+s}- L_{r+s} + \Lambda^{(2)}_{r+s} \; ,
\end{eqnarray}
where
\begin{equation}\label{lambda}
\Lambda^{(2)}_n = -\frac{18}{c}\sum_m U_{n-m}U_m \quad .
\end{equation}
The indices $m , n$ and $r, s$ have integer and half integer
values, respectively, and run from $-\infty$ to $+\infty$.

The algebra \p{1} is nonlinear. A useful idea \cite{IKP} is  to
consider the composite $\Lambda^{(2)}_n$ as a
new higher-spin generator. Commutators of $\Lambda^{(2)}_n$ with other
generators give us new composites, which we will treat as additional
generators and so on.
The $W_3^{(2)\infty}$ algebra is then defined as the  set of generators
$L_n,G^{\pm}_r,U_n,\Lambda^{(2)}_n$ and all higher-spin composites
$\Lambda^{(s)}_n\quad (s\geq 3)$ which appear in the commutators
of the lower-spin generators.
Thus, the $W_3^{(2)\infty}$ algebra containes an infinite set
of generators with increasing conformal weights. The commutation
relations involving the higher-spin generators $\Lambda^{(s)}$ can be
computed using the basic relations \p{1}-\p{lambda}.

Following Refs.\cite{{IKP},{IKM}}, in constructing the nonlinear realization
of $W_3^{(2)\infty}$ we deal with
the following subalgebra of $W_3^{(2)\infty}$, rather than
the whole algebra:\footnote{In what follows
we deal only with this subalgebra, so we use for it the same
notation $W_3^{(2)\infty}$.}
\begin{equation} \label{2}
L_n, \;  G_{n+\frac{1}{2}}^{\pm}, \; U_{n+1}, \;  \Lambda^{(s)}_{n+2-s}, \;
(n\geq -1).
\end{equation}
It is interesting to notice
that the wedge algebra $W_{\wedge}$ defined as the
subalgebra of \p{2} with indices varying from $-(s-1)$ to $(s-1)$
for each spin $s$ contains, in close analogy with the
$W_3$ case\cite{IKP},
the $sl(3,R)$ factor algebra spanned by the generators
\begin{equation}\label{sl3}
\left\{ L_{-1},L_0,L_{1},G_{-1/2}^{\pm},G_{1/2}^{\pm},U_0 \right\}
    \sim sl(3,R) \; .
\end{equation}
We will explore this factor algebra in the next Section, where we consider
the Toda-type nonlinear realization of $W_3^{(2)\infty}$.

Any nonlinear realization is specified by the choice of the stability
subgroup $H$ or, equivalently, its stability subalgebra $\cal H$.
There are many
different subalgebras in the $W_3^{(2)\infty}$ algebra. Next, let us
discuss the main peculiarities of the subalgebras we are looking for.

First of all, as in Refs.\cite{{IKP},{IKM}}, we always put all higher-spin
generators in the stability subalgebra. Let us remark that all higher-spin
generators, by construction, are at least quadratic in the basic
generators $L,G^{\pm},U$. Therefore, under the classical commutation relations
\p{1} they form themselves a subalgebra $\cal HS$.

Secondly, let us remind that in the nonlinear realizations approach
\cite{coset}, the parameters associated with the coset generators
have the meaning of coordinates and/or fields.
Among the remaining generators $U_n,L_n,G_r^{\pm}$ in the
$W_3^{(2)\infty}$ algebra \p{2} three possess
a negative dimension: $L_{-1} (cm^{-1}), G_{-1/2}^{\pm} (cm^{-1/2})$.
Following the $W_3$ case \cite{{IKP},{IKM}}, it is
natural to associate the coordinate $x$
with the generator $L_{-1}$, so we will keep this
generator in the coset. As for the $G_{-1/2}^{\pm}$
generators, the dimension of the corresponding coset parameters $(cm^{1/2})$
is unsuitable  for treating them as fields. On the other hand, we cannot put
both $G_{-1/2}^{+}$ and $G_{-1/2}^{-}$
in the stability subgroup, because their commutator
yields $L_{-1}$. So, when
both $G_{-1/2}^{+}$ and $G_{-1/2}^{-}$
are present in the coset (and the stability subgroup coincides
with ${\cal HS}$ ), we have to introduce two additional coordinates
$t^{\pm}$ with dimension $(cm^{1/2})$.
We postpone the discussion of this 3-dimensional case for the future,
limiting our consideration here
to the following three possibilities ${\cal H},
{\cal H}_1,{\cal H}_2$ for the stability subgroup:
\begin{eqnarray}
{\cal H} & = & \left\{ G_{-1/2}^{-},\; L_0,\; U_0,\;
G_{1/2}^{\pm},\; L_1,\;\cal{HS}\right\}, \label{H} \\
{\cal H}_1& = & \left\{ G_{-1/2}^{-}, \; L_0\;, U_0,\;\cal{HS}\right\},
                  \label{H1}\\
{\cal H}_2& = & \left\{ G_{-1/2}^{-}, \; \cal{HS}\right\}. \label{H2}
\end{eqnarray}
In all these cases
we associate a "time" coordinate $t$ with the linear combination
$G_{-1/2}^{+}+G_{-1/2}^{-}$, keeping the latter in the coset. All other coset
generators $L_n,G^{\pm}_r,U_m$ for the stability subgroups
${\cal H},{\cal H}_1,{\cal H}_2$ have growing
positive dimensions, so the corresponding parameters can be associated with
fields that depend on the coordinates $x$ and $t$.

Keeping these facts in the mind, let us turn to the construction of
the nonlinear realization of the $W_3^{(2)\infty}$ algebra \p{1}-\p{2}.

The minimal coset space corresponds to the maximal
subgroup $\cal H$ \p{H} and contains the generators
$\left\{ L_{-1}, G_{-1/2}^{+}+G_{-1/2}^{-}, U_n, G_{n+1/2}^{\pm},
L_{n+1} \right\}$ with $n\geq 1$. It can be parametrized
in the following form:
\begin{equation} \label{g}
g = e^{xL_{-1}}e^{ t(G_{-1/2}^{+}+G_{-1/2}^{-})}
\left( \prod_{n\geq 3}
e^{\xi^{\pm}_{n-1/2}G_{n-1/2}^{\pm}}
e^{u_nL_n}e^{\phi_{n-1}U_{n-1}}\right)
e^{\phi_1 U_1}
e^{\xi^{+}G_{3/2}^{+}+\xi^{-}G_{3/2}^{-}}
e^{u_2 L_2} .
\end{equation}
The parameters $u_n,\phi_n,\xi^{\pm}_r$ depend on $x$ and $t$.
As usual in the nonlinear realization approach \cite{coset}, the group
$W_3^{(2)\infty}$ acts on the coset \p{g} as the left multiplication
\begin{equation}\label{trans}
g_0 \cdot g = g' \cdot h \quad ,
\end{equation}
where $g_0$ is an arbitrary element of $W_3^{(2)\infty}$ group and
$h$ belongs to the stability group $H$. This provides the realization
of  $W_3^{(2)\infty}$ in the space spanned by the coordinates $x,t$
and the infinite number of fields $u_n,\phi_n,\xi^{\pm}_r$.

To reduce the number of fields we need to invoke the inverse Higgs
effect \cite{IO} by putting some appropriate covariant
constraints on the Cartan forms, which are defined in the standard way
\begin{equation}\label{forms}
\Omega = g^{-1}dg = \sum \omega_i A_i  + \sum {\tilde\omega}_k V_k \quad .
\end{equation}
Here we denote all generators from the stability subalgebra ${\cal H}$ \p{H}
and coset \p{g} as $V_k$ and $A_i$ respectively. It is well known \cite{coset},
that the forms $\omega_i$ associated with the coset generators $A_i$
transform homogeneously under the $W_3^{(2)\infty}$ transformations \p{trans}.
So, we may put some of them to zero. These constraints will be
covariant if the remaining non-zero forms belong to the generators spanning
a subalgebra together with all stability subalgebra generators $V_k$.
In the case at hand, we may extend the stability subalgebra $\cal H$ \p{H}
only by the generators $L_1$ and $G^{+}_{-1/2}+G^{-}_{-1/2}$. Thus, the
corresponding unique set of covariant constraints reads
\begin{equation}\label{constr}
\omega_n^L=\omega_{n-\frac{1}{2}}^{\pm}=\omega_{n-1}^U=0 \;,\;\mbox{with }
  n\geq 2.
\end{equation}
Let us present explicitly the first forms and the corresponding equations
\begin{equation}\label{eq1}
\omega_1^U=d\phi_1-3(\xi^{+}+\xi^{-})dt-2\phi_2dx=0:\left\{\begin{array}{l}
\phi_1'-2\phi_2=0,\\
{\dot\phi}_1-3(\xi^{+}+\xi^{-})=0,
\end{array}\right.
\end{equation}
\begin{equation}\label{eq2}
\omega_{\frac{3}{2}}^{+}=
d\xi^{+} -\xi^{+}\phi_1dx+(\frac{1}{2}
\phi_1^2-\phi_2 -\frac{3}{2}u_2)dt-3\xi_{\frac{5}{2}}^{+}dx=
0:\left\{\begin{array}{l}
{\xi^{+}}'
-\xi^{+}\phi_1-3\xi_{\frac{5}{2}}^{+}=0,\\
{\dot\xi}^{+}+\frac{1}{2}\phi_1^2-\phi_2-\frac{3}{2}u_2=0,
\end{array}\right.
\end{equation}
\begin{equation}\label{eq3}
\omega_{\frac{3}{2}}^{-}=
d\xi^{-} +\xi^{-}\phi_1dx+(\frac{1}{2}
\phi_1^2+\phi_2 -\frac{3}{2}u_2)dt-3\xi_{\frac{5}{2}}^{-}dx =
0:\left\{\begin{array}{l}
{\xi^{-}}'
+\xi^{-}\phi_1-3\xi_{\frac{5}{2}}^{-}=0,\\
{\dot\xi}^{-}+\frac{1}{2}\phi_1^2+\phi_2-\frac{3}{2}u_2=0,
\end{array}\right.
\end{equation}
\begin{equation}\label{eq4}
\omega_2^L=
du_2+(\xi_{\frac{5}{2}}^{+}-\xi_{\frac{5}{2}}^{-} +
\xi^{+}\phi_1+ \xi^{-}\phi_1)dt-4u_3dx =
0:\left\{\begin{array}{l}
u'_2-4u_3=0,\\
{\dot u}_2+\phi_1(\xi^{+}+\xi^{-})+
\xi_{\frac{5}{2}}^{+}-\xi_{\frac{5}{2}}^{-}=0,
\end{array}\right.
\end{equation}
where we denote by prime and dot the $x$ and $t$ derivatives, respectively.

The first equations in \p{eq1}-\p{eq4} express the higher coset fields
in terms of the lower ones $u_2,\phi_1,\xi^{\pm}$:
\begin{eqnarray}\label{IH}
 & & u_3=\frac{1}{4}u'_2 \; , \;\phi_2=\frac{1}{2}\phi'_1 \; , \nn
 & & \xi_{\frac{5}{2}}^{+}=\frac{1}{3} ( {\xi^{+}}'-\xi^{+}\phi_1 ) \; , \;
     \xi_{\frac{5}{2}}^{-}=\frac{1}{3} ( {\xi^{-}}'+\xi^{-}\phi_1 )\; .
\end{eqnarray}
The meaning of these equations is simple: they claim that the transformation
properties of the fields $u_3,\phi_2,\xi_{5/2}^{\pm}$ under the
$W_3^{(2)\infty}$
group \p{trans} coincide with the transformations of the right hand side
expressions in Eqs. \p{IH}. Thus, in all calculations we may ignore the
fields  $u_3,\phi_2,\xi_{5/2}^{\pm}$ substituting their expressions in terms
of the essential ones $u_2,\phi_1,\xi^{\pm}$.

The second equations in \p{eq1}-\p{eq4} give the equations of motion for
the essential fields (after using \p{IH}):
\begin{eqnarray}\label{EOM}
{\dot\phi}_1 &= & 3(\xi^{+}+\xi^{-}) \; , \nn
{\dot\xi}^{+}& = & \frac{3}{2}u_2+\frac{1}{2}\phi'_1-\frac{1}{2}\phi_1^2 \;,\nn
{\dot\xi}^{-}& = & \frac{3}{2}u_2-\frac{1}{2}\phi'_1-\frac{1}{2}\phi_1^2 \;,\nn
{\dot u}_2 & = & \frac{1}{3}({\xi^{-}}'-{\xi^{+}}')-\frac{2}{3}\phi_1
           (\xi^{+}+\xi^{-}) \; .
\end{eqnarray}
The $x$-Boussinesq equation
\begin{equation}\label{Bous}
\phi_1''-\frac{1}{3}\frac{\partial^2}{\partial t^2}
 \left( \frac{\partial^2}{\partial t^2}{\phi}_1 +4\phi_1^2\right)  =0
\end{equation}
arises as a consistency condition for this system.

It can be checked that the remaining infinitely many equations in \p{constr}
express the higher fields $u_n,\phi_n,\xi^{\pm}_r$
in terms of $u_2,\phi_1,\xi^{\pm}$, and do not put any additional dynamical
constraints on the essential fields, besides \p{EOM}.
Thus, we deduced the $x$-Boussinesq equation in a purely
geometrical way, starting from
the nonlinear realization of $W_3^{(2)\infty}$ together with the constraints
\p{constr}.

What happens if we consider the narrow stability subalgebra ${\cal H}_1$?
Let us take the following parametrizations for the coset:
\begin{equation}
g_1  =  g e^{\eta^{+}G^{+}_{1/2}+\eta^{-}G^{-}_{1/2}}e^{u_1L_1} . \label{hh1}
\end{equation}
It can be checked explicitly, that using the previous constraints \p{constr}
gives the same equations \p{eq1}-\p{eq4}.
However, owing to the presence of new generators in the coset \p{hh1},
we could extend our constraints \p{constr} by the following
additional ones:
\begin{equation}\label{constr1}
\omega_1^L=\omega^{\pm}_{\frac{1}{2}} =0 \;.
\end{equation}
Without going into the details, let us write
the new equations,  besides \p{IH}, \p{EOM}
\begin{eqnarray}
\xi^{+}&=&\frac{1}{2} \left(
 {\eta^{+}}'+\frac{1}{2}\eta^{+}\eta^{+}\eta^{-}+
      (u_1-\phi_1 )\eta^{+} \right)  \; , \nn
\xi^{-}&=&\frac{1}{2}\left(
 {\eta^{-}}'-\frac{1}{2}\eta^{-}\eta^{-}\eta^{+}+
      (u_1+\phi_1 )\eta^{-} \right)  \; , \nn
u_2 & = & \frac{1}{3}\left( u'_1+u_1^2+\frac{3}{2}(\eta^{+}{\eta^{-}}'-
     \eta^{-}{\eta^{+}}' )-\eta^{+}\eta^{-}
   (\phi_1-\frac{3}{4}\eta^{+}\eta^{-}) \right)\; , \label{eh1}
\end{eqnarray}
and
\begin{eqnarray}\label{eeh1}
\dot{u}_1 & = & \frac{1}{2}\left(
 \eta^{-}{\dot\eta}^{+}-\eta^{+}{\dot\eta}^{-}
  +{\eta^{-}}'-{\eta^{+}}'+u_1(\eta^{+}-\eta^{-})-
  (\eta^{+}+\eta^{-})(\phi_1+\frac{3}{2}\eta^{+}\eta^{-})\right) \;, \nn
\dot{\eta}^{+}& = & \eta^{+}\left( \eta^{+}+\frac{1}{2}\eta^{-}\right)+
         u_1+\phi_1 \; , \nn
\dot{\eta}^{-}& = & -\eta^{-}\left( \eta^{-}+\frac{1}{2}\eta^{+}\right)+
         u_1-\phi_1 \; .
\end{eqnarray}
The equations \p{eh1} express the fields $u_2,\xi^{\pm}$ with spins $2,3/2$,
respectively, in terms of the new fields $u_1,\phi_1,\eta^{\pm}$
with spins $1,1,1/2$, which obey the dynamical equations \p{eeh1}.
Thus, we recognize the equations \p{eh1}-\p{eeh1} as Miura transformations
for the system \p{EOM}.

Analogous calculations for the stability subalgebra ${\cal H}_2$ with
the following parametrization of the coset $g_2$ :
\begin{equation}
g_2  =  g_1 e^{u_0L_0}e^{\phi_0U_0} \; , \label{hh2}
\end{equation}
and the additional constraints:
\begin{equation}\label{constr2}
\omega_0^L=\omega_0^U =0 \;
\end{equation}
give the next Miura transformations
\begin{equation}\label{le1}
\phi_1  =  \phi'_0+\frac{3}{2}\eta^{+}\eta^{-} \; , \;
u_1  =  \frac{1}{2}u'_0
\end{equation}
and
\begin{equation}\label{le2}
{\dot\phi}_0  =  \frac{3}{2}(\eta^{+} + \eta^{-} ) \; , \;
{\dot u}_0  =  \eta^{-}-\eta^{+} .
\end{equation}

Thus, passing from the stability subgroup $\cal H$ to ${\cal H}_1$ and
${\cal H}_2$ and putting the corresponding constraints on the Cartan forms,
we obtain the Miura transformations, expressing the higher-spin
fields in terms of lower-spin ones, as well as the equations of
motion for the latter. It can be checked, that all these systems
\p{EOM},\p{eh1},\p{eeh1},\p{le1},\p{le2} are self-consistent, i.e.
starting from the lowest spin fields $u_0,\phi_0,\eta^{\pm}$ and their
equations of motion one could recover the equations of motion for
the higher-spin fields $u_1,\phi_1,\eta^{\pm}$ and $u_2,\phi_1,\xi^{\pm}$
defined through the Miura transformations \p{le1} and \p{eh1}.

Let us finish this Section with some comments.

First, it is clear that due to presence of the generator $G^{-}_{-1/2}$ in
all stability subalgebras ${\cal H}$, ${\cal H}_1$
and ${\cal H}_2$ \p{H}-\p{H2}, we are
free to associate the $t$-coordinate with the  generator
$G_{-1/2}^{+}+\beta G_{-1/2}^{-}$:
\begin{equation} \label{gbeta}
{\tilde g} = e^{xL_{-1}}e^{ t(G_{-1/2}^{+}+\beta G_{-1/2}^{-})}\ldots \quad ,
\end{equation}
where $\beta$ is an arbitrary parameter.
All values of $\beta$ are equivalent up to the corresponding renormalizations,
except the case $\beta = 0$. In this case we get the following
equations of motion instead of \p{EOM}:
\begin{eqnarray}\label{EOM1}
{\dot{\widetilde \phi}}_1 &= & 3{\widetilde\xi}^{-} \; , \nn
{\dot{\widetilde{\xi^{+} }}}& = & \frac{3}{2}{\widetilde u}_2+
 \frac{1}{2}{\widetilde\phi}'_1-\frac{1}{2}{\widetilde\phi}_1^2 \;,\nn
{\dot{\widetilde{\xi^{-} }}}& = & 0 \;,\nn
{\dot {\widetilde u}}_2 & = & \frac{1}{3}{\widetilde{\xi^{-}}}'-
 \frac{2}{3}{\widetilde\phi}_1{\widetilde\xi}^{-} \; .
\end{eqnarray}
Hence the field ${\widetilde\phi}_1$ obeys the free equation of motion
\begin{equation}
\ddot{{\widetilde\phi}}_1 =0 \;
\end{equation}
rather than the $x$-Boussinesq one \p{Bous}. Close inspection shows that the
equations \p{EOM1} can be explicitly integrated. But in the nonlinear
realization framework both choices of the coset element \p{g} and \p{gbeta}
are equivalent and are connected through a right transformation by the
stability subgroup
\begin{equation}
g={\tilde g} \; h \; , \; h\subset H \; .
\end{equation}
In principle, one could find this transformation explicitly in a purely
algebraic way, although of course in practice the calculations will
rapidly become unmanageable. Thus, the transformation (2.30) to
this new coset parametrization
\p{gbeta} gives the action-angle variables which allow to rewrite
the
equations of motion \p{EOM} and the $x$-Boussinesq equation \p{Bous}
in the trivial form (2.28) and (2.29), respectively.

Second, in the proposed approach the invariance of the equations \p{EOM},
\p{eeh1},\p{le2}
with respect to $W_3^{(2)\infty}$ transformations is evident due to the
covariance of the constraints \p{constr},\p{constr1},\p{constr2}. The explicit
transformation laws for the coordinates and fields can be easily found
from the general formula \p{trans}.

Finally, let us remark that in accordance with the inhomogeneous transformation
laws of the Goldstone fields we can associate the fields
$\phi_1,\xi^{\pm},u_2$ with the currents $U,G^{\pm},T$ that generate the
$W_3^{(2)}$ algebra \cite{Polyakov} under appropriate Poisson brackets.
Then, the Miura maps \p{eh1},\p{le1}
give the free fields representation for the currents, in close analogy
with the $W_3$ case \cite{IKM}.

\setcounter{equation}0
\section{$SL(3,R)$-Toda lattice}

We have seen in the previous Section how the $x$-Boussinesq equation is
connected with the nonlinear realization of the algebra $W_3^{(2)\infty }$
\p{1}. All fields in the coset space are  functions of the
two coordinates $x$ and $t$, whose dimensionalities are, respectively,
$1$ and $\frac{1}{2}$.
In this Section we show that 2D $sl_3$ Toda equations of motion
also result from a particular coset realization of $W_3^{(2)\infty}$.

In order to have manifest 2D Lorentz symmetry, we start from the product
of two commuting copies $W_3^{(2)(\pm)}$ of the algebra. As a next step
we need to specify the stability subalgebra ${\cal H}$.
First of all, we put, as in the previous Section, all higher-spin
composite generators into the stability subalgebra and neglect them
in what follows.
Secondly, in order to have a real two-dimensional space with coordinates
of the same
dimensionality, we need to keep in the coset only two generators with
negative dimension and associate with them the coordinates $x^\pm$.
Let us remind, that six of the generators  of the two copies of
$W_3^{(2)\infty}$ algebras \p{2} possess
a negative dimension, i.e. $L_{-1}^{(\pm )}, G_{-1/2}^{\pm (\pm )} $.
The only possibility to have two commuting $x^\pm $ coordinates is to
associate them with the generators $ G_{-1/2}^{+(\pm)}+ G_{-1/2}^{-(\pm)}$.
Finally, we put the generator of the Lorentz rotation $L^{(+)}_0-L^{(-)}_0$
in the stability subalgebra, with the Liouville field $u(x)$ as the parameter
corresponding to the coset generator $L^{(+)}_0+L^{(-)}_0$.
But we need also to
introduce the second Toda field $\phi (x)$ as a coset parameter.
The only appropriate generators are $U_0^{\pm }$, so we include a linear
combination of them $U^{(+)}_0+U^{(-)}_0$ into the coset.

All these requirement are satisfied from the two-parameter family of
the stability subgroups generated by
\begin{eqnarray} \label{TodaH}
{\cal H} & = & \left\{ L_{-1}^{(\pm)}-m_1m_2L_{1}^{(\mp)}\; , \;
G_{-1/2}^{+(+)}+m_1G_{1/2}^{-(-)}\;,\;G_{-1/2}^{-(-)}+m_2G_{1/2}^{+(+)},
        \right. \nn
 & & \left. L_0^{(+)}-L_0^{(-)} \; , \; U_0^{(+)}-U_0^{(-)}\; , \;
    \cal{HS}^{(\pm)}\right\}.
\end{eqnarray}
Let us note that the combinations of generators $L,G$ and $U$ in \p{TodaH}
form, in close analogy with the $W_3$ case \cite{IKP}, the Borel subalgebra
of the diagonal $sl(3,R)$ in the sum of the two commuting factor-algebras
$sl(3,R)^{\pm}$ \p{sl3} in $W_3^{(2)\infty (+)}$ and $W_3^{(2)\infty (-)}$.
We parametrize the resulting coset space as follows:
\begin{eqnarray}\label{Tg}
g & =&  e^{ x^{+}(G_{-1/2}^{+(+)}+G_{-1/2}^{-(+)})}
        e^{ x^{-}(G_{-1/2}^{+(-)}+G_{-1/2}^{-(-)})}
\prod\limits_{n=1}^{\infty}\left(
e^{\xi^{+(\pm)}_{n-1/2}G_{n-1/2}^{+(\pm)}+
\xi^{-(\pm)}_{n-1/2}G_{n-1/2}^{-(\pm)}}
e^{u_n^{(\pm)}L_n^{(\pm)}+\phi_n^{(\pm)}U_n^{(\pm)}}\right) \nn
 & & e^{u(L_0^{(+)}+L_0^{(-)})} e^{\phi(U_0^{+}+U_0^{-})}.
\end{eqnarray}

Next, we show that after imposing appropriate
covariant constraints one can
express the higher fields in terms of $u(x)$ and $\phi (x)$ which
obey the $sl_3$ Toda equations. Defining Cartan forms in the standard
way \p{forms}, one constrains them as follows:
\begin{equation}\label{TodaCon}
g^{-1}dg = \sum \omega^{L^{(\pm )}}_nL_n^{(\pm )} +
                 \sum \omega^{G^{\pm (\pm )}}_r G^{\pm (\pm )}_r +
                 \sum \omega^{U^{(\pm )}}_n U^{(\pm )}_n +\ldots =
       {\tilde g}^{-1}d{\tilde g} \in {\widetilde{\cal H}} \; ,
\end{equation}
where ${\widetilde{\cal H}}$ is some subalgebra containing the stability
subalgebra $\cal H$ \p{TodaH}. In other words, we put all forms to zero
except those which belong to the generators of the subalgebra
${\widetilde{\cal H}}$. It is natural to choose the following subalgebra
${\widetilde{\cal H}}$:
\begin{eqnarray} \label{sl}
{\widetilde{\cal H}} &= &\left\{ L_{-}=L_{-1}^{-}-m_1m_2L_1^{+},\;\;\;\;\;
L_{+}=L_{-1}^{+}-m_1m_2L_1^{-}, \right. \nn
&&G^{--}=G_{-\frac{1}{2}}^{-(-)}+m_2G_{\frac{1}{2}}^{+(+)},\;\;
G^{-+}=G_{-\frac{1}{2}}^{-(+)}+m_2G_{\frac{1}{2}}^{+(-)}, \nn
&&G^{+-}=G_{-\frac{1}{2}}^{+(-)}+m_1G_{\frac{1}{2}}^{-(+)},\;  \;
G^{++}=G_{-\frac{1}{2}}^{+(+)}+m_1G_{\frac{1}{2}}^{-(-)},\nn
&&\left. L=L_0^{+}-L_0^{-},\;\; U=U_0^{+}-U_0^{-}, {\cal HS}^{\pm}
         \right\} \;.
\end{eqnarray}
This algebra is an extension of the stability subalgebra $\cal H$ \p{TodaH},
obtained by including the generators $G^{+-}$ and $G^{-+}$.
Let us stress that all higher-spin generators still form an ideal
in \p{sl} and the factor-algebra of ${\widetilde{\cal H}}$ over this ideal
is the diagonal $sl(3,R)$. The parameters $m_1,m_2$ correspond to
the freedom  in extracting the diagonal $sl(3,R)$ from
$sl(3,R)^{(+)}\times sl(3,R)^{(-)}$.

Each of the infinite number of equations \p{TodaCon} gives rise to two
equations for the $dx^+$ and $dx^-$ projection of forms. We quote the
first few equations explicitely
\begin{eqnarray}\label{T1}
&& 2\partial_{+}u+\xi^{+(+)}-\xi^{-(+)}=0 \; ,\;
2\partial_{-}u+\xi^{+(-)}-\xi^{-(-)}=0,         \nn
&& 2\partial_{+}\phi-\frac{3}{2}(\xi^{+(+)}+\xi^{-(+)})=0 \; ,\;
2\partial_{-}\phi-\frac{3}{2}(\xi^{+(-)}+
\xi^{-(-)})=0, \nn
& & \partial_{+}\xi^{+(+)}-\phi_1^{(+)}-u_1^{(+)}-
\frac{1}{2}\xi^{+(+)}\xi^{-(+)}-
 \xi^{+(+)}\xi^{+(+)}=0 \; , \nn
&& \partial_{+}\xi^{-(+)}+\phi_1^{(+)}-u_1^{(+)}+
 \frac{1}{2}\xi^{+(+)}\xi^{-(+)}+
 \xi^{-(+)}\xi^{-(+)}=0 \; , \nn
&& \partial_{-}\xi^{+(-)}-\phi_1^{(-)}-u_1^{(-)}-
      \frac{1}{2}\xi^{+(-)}\xi^{-(-)}-
      \xi^{+(-)}\xi^{+(-)}=0 \; , \nn
&& \partial_{-}\xi^{-(-)}+\phi_1^{(-)}-u_1^{(-)}+
\frac{1}{2}\xi^{+(-)}\xi^{-(-)}+
\xi^{-(-)}\xi^{-(-)}=0
\end{eqnarray}
and
\begin{eqnarray}\label{T2}
& & \partial_{-}\xi^{+(+)}-m_{-}e^{-u+2\phi}=0 \; , \;
 \partial_{-}\xi^{-(+)}-m_{+}e^{-u-2\phi}=0, \nn
& & \partial_{+}\xi^{+(-)}-m_{-}e^{-u+2\phi}=0 \; , \;
\partial_{+}\xi^{-(-)}-m_{+}e^{-u-2\phi}=0\;.
\end{eqnarray}
We use the notations $\partial_{\pm}=\partial/\partial x^{\pm}$.

The equations \p{T1} are purely algebraic and allow to express the higher-spin
coset fields in terms of $u(x)$ and $\phi (x)$. Let us stress that all
coset fields can be eliminated in this manner, e.g.:
\begin{eqnarray}
& & \xi^{-(+)} = \partial_{+}u+\frac{2}{3}\partial_{+}\phi \; , \;
  \xi^{+(+)} = -\partial_{+}u+\frac{2}{3}\partial_{+}\phi \; , \nn
& & \xi^{-(-)} = \partial_{-}u+\frac{2}{3}\partial_{-}\phi \; , \;
  \xi^{+(-)} = -\partial_{-}u+\frac{2}{3}\partial_{-}\phi \; \mbox{etc.}
\end{eqnarray}
The remaining equations \p{T2} constrain $u(x), \phi (x)$ to obey the $sl_3$
Toda equation
\begin{eqnarray}\label{lasteq}
\partial_{+-}u&=&\frac{1}{2}\left(m_{1}e^{-u-2\phi}-
m_{2}e^{-u+2\phi}\right),\nn
\partial_{+-}\phi&=&\frac{3}{4}\left(m_{1}e^{-u-2\phi}+
m_{2}e^{-u+2\phi}\right).
\end{eqnarray}

Thus, we succeeded in constructing the $sl_3$ Toda lattice equations
starting from a purely geometric coset realization of
$W_3^{(2)\infty (+)}\times W_3^{(2)\infty (-)}$ symmetry.
It is worth noticing that the dimensionality of the coordinates in \p{lasteq}
is $cm^{1/2}$ in contrast with the ordinary 2D Minkowski space coordinates.
However, we are free to renormalize this coordinates at our wishes.
Finally, let us stress that the $sl_3$ Toda lattice equations are
{\it by construction} invariant with respect to the
$W_3^{(2)+}\times W_3^{(2)-}$ symmetry realized as left shifts of the
coset \p{Tg}.

\setcounter{equation}0
\section{Conclusion}

The abovecited recent literature elucidates many facets of the
intimate relationship between the $W$-type algebras, on the one
hand, and the
conformal field theory and integrable systems in 1+1 dimensions,
on the other hand. Such close connection has far-reaching implications
for two-dimensional gravity and string theories.

In this letter we have demonstrated, in the simplest example of the
$W_3^{(2)}$ symmetry, that the covariant reduction approach can be applied
also to the nonlinear algebras with non-canonical spin generators.
We have shown that the Boussinesq equation in $x$-evolution together with
the Miura maps for it and the
$sl_3$ Toda system have a clear geometric origin:
they  can be derived from a coset realization of the linear algebra
$W_3^{(2)\infty }$.

Actually, this work is part of a more ambitious
project whose goal is to obtain (and maybe classify) all 2D integrable
systems from the nonlinear realizations of some infinite-dimensional
(nonlinear) algebras. The advantage of such approach, apart from its
clear geometric meaning, is that it yields directly the equations of
motion, the Miura maps, as well as the
transformation properties of the fields and
coordinates for the given algebra. In the framework of the
proposed approach, one of the most intriguing questions
is how to establish a relation between $W_3$ and
$W_3^{(2)}$ algebras which give rise, in essence, to the same
integrable systems. This issue is currently
under consideration and further results will be reported elsewhere.

\def\thesection { }
\section{Acknowledgements}
It is pleasure to thank E. Ivanov and A. Sorin for their
interest in this work and very clarifying discussions.
Many thanks go to JINR - Bogoliubov Theoretical Laboratory and
Prof. E. Shabalin at ITEP for the warm hospitality extended to S.B. when this
work was undertaken.

\end{document}